\documentclass[11pt,twoside]{article}
\usepackage{amssymb,amsmath}

\newcounter{proposition}

\newcommand{\nothing}[1]{}

\newcommand{\beq}[1]{\begin{equation}\label{#1}}
	\newcommand{\eeq}{\end{equation}}

\newcommand{\bmu}[1]{\begin{multline}\label{#1}}
	\newcommand{\emu}{\end{multline}}

\newcommand{\eq}{\triangleq}

\newcommand{\x}{{\textbf{\textit{x}}}}

\renewcommand{\u}{{\bf u}}
\renewcommand{\v}{{\bf v}}
\renewcommand{\S}{{\cal S}}

\renewcommand{\S}{{\mathcal{S}}}

\renewcommand{\chi}{\upsilon}

\renewcommand{\[}{\left[}

\newcommand{\lev}{\left\lceil}
\newcommand{\riv}{\right\rceil}
\renewcommand{\le}{\leqslant}

\usepackage{fancyhead}
\renewcommand{\sectionmark}[1]{}
\renewcommand{\subsectionmark}[1]{}
\begin{document}

	\vspace*{5mm}
	
	\noindent
	%%--The title should be placed here
	\textbf{\LARGE On a  Hypergraph Approach to Multistage
		Group Testing  Problems 
		%%--If you do not use any support, just comment the next line or delete it.
		\footnote{The research is supported in part by the Russian Foundation for Basic Research under
			Grant No. 16-01-00440.}}
	\thispagestyle{fancyplain} \setlength\partopsep {0pt} \flushbottom
	\date{}
	
	\vspace*{5mm}
	\noindent
	%%--The name(s) of the author(s), their e-mails and addresses should be placed here
	\textsc{A. G. D'yachkov} \hfill \texttt{agd-msu@yandex.ru} \\
	{\small Lomonosov Moscow State University, Moscow, Russia} \\[3pt]
	\textsc{I.V. Vorobyev} \hfill \texttt{vorobyev.i.v@yandex.ru} \\
	{\small Lomonosov Moscow State University, Moscow, Russia} \\[3pt]
	\textsc{N.A. Polyanskii} \hfill \texttt{nikitapolyansky@gmail.com} \\
	{\small Lomonosov Moscow State University, Moscow, Russia} \\[3pt]
	\textsc{V.Yu. Shchukin} \hfill \texttt{vpike@mail.ru} \\
	{\small Lomonosov Moscow State University, Moscow, Russia} \\[3pt]
	
	\medskip
	
	\begin{center}
		\parbox{11,8cm}{\footnotesize
			%%--The abstract goes here.
			\textbf{Abstract.} 
			Group testing is a well known search problem that consists in detecting up to $s$
			defective elements of the set $[t]=\{1,\ldots,t\}$ by carrying out tests on properly chosen subsets
			of $[t]$. In classical group testing the goal is to find all defective elements by
			using the minimal possible number of tests. In this paper we consider multistage group testing. We propose a general idea how to use a hypergraph approach to searching defects. For the case $s=2$, we design an explicit construction, which makes use of $2\log_2t(1+o(1))$ tests in the worst case  and consists of $4$ stages.}
		
	\end{center}
	
	\baselineskip=0.9\normalbaselineskip
	
	\section{Introduction}
	\lhead{}
	\rhead{}
	\chead[\fancyplain{}{\small\sl\leftmark}]{\fancyplain{}{\small\sl\leftmark}}
	\cfoot{}
	\markboth{\hspace{-0.2cm}Fifteenth International Workshop on Algebraic and Combinatorial Coding Theory\\ June 18-24, 2016, Albena, Bulgaria \hfill pp. 145--150}{}
	\setcounter{page}{145}
	
	Group testing is a  very natural combinatorial problem that consists in detecting up to $s$
	defective elements of the set of objects $[t]=\{1,\ldots,t\}$ by carrying out tests on properly chosen subsets (pools)
	of $[t]$. The test outcome is positive if the tested pool contains one
	or more defective elements; otherwise, it is negative.

	There are two general types of algorithms. In \textit{adaptive}
	group testing, at each step the algorithm decides which group to test by observing
	the responses of the previous tests. In \textit{non-adaptive} algorithm, all tests are carried out in parallel. There is a compromise algorithm between these two types, which is called a \textit{multistage} algorithm. For the multistage algorithm all tests are divided into $p$ sequential stages. The tests inside the same stage are performed simultaneously. The tests of the next stages may depend on the responses of the previous. In this context, a non-adaptive group testing algorithm is reffered to as a  one stage algorithm.
	\subsection{Previous results}
	We refer the reader to the monograph  \cite{DH} for a survey on group testing and its applications. In spite of the fact that the problem of estimating the minimum \textit{average} (the set of defects is chosen randomly) number of tests has been investigated in many papers (for instance, see \cite{ds13, mt11}), in the given paper we concentrate our attention only on the minimal number of test in the \textit{worst case}. 
	
	In 1982 \cite{dr82}, Dyachkov and Rykov proved that at least 
	$$
	\frac{s^2}{2\log_2(e(s+1)/2)}\log_2 t(1+o(1))
	$$ 
	tests are needed for non-adaptive group testing algorithm. 
	
	If the number of stages is $2$, then it was proved that $O(s \log_2 t)$ tests are already sufficient. It was shown by studying random coding bound for disjunctive list-decoding codes \cite{r90,d03} and  selectors \cite{BGV}. The recent work \cite{DVPS} has significantly improved the constant factor in the main term of number of tests for two stage group testing procedures. In particular, if $s\to\infty$, then 
	$$\frac{se}{\log_2e}\log_2t (1+o(1))$$
	tests are enough for two stage group testing.  
	
	As for adaptive strategies, there exist such ones that attain the information theory lower bound $s \log_2t (1+o(1))$. However, for $s>1$ the number of stages in well-known optimal strategies  is a function of $t$, and grows to infinity as $t\to\infty$.

	\subsection{Summary of the results}
	\headrulewidth 0pt
		\lhead[\fancyplain{}{\thepage}]{\fancyplain{}{\rightmark}}
		\rhead[\fancyplain{}{\leftmark}]{\fancyplain{}{\thepage}}
		\lfoot{}
		\rfoot{}
		\chead{}
		\cfoot{}
		\markboth{\textsl{ACCT2016}}{\textsl{D'yachkov, Vorobyev, Polyanskii, Shchukin}}
	In the given article we present some explicit algorithms, in which we make a restriction on the number of stages. It will be a function of $s$. We briefly give necessary notations in section \ref{Pre}. Then, in section~\ref{Hyp}, we present a general idea of searching defects using a hypergraph approach.  In section~\ref{Search2}, we describe a $4$-stage group testing  strategy, which detects $2$ defects and  uses the asymptotically optimal number of tests  $2\log_2t(1+o(1))$. As far as we know the best result for such a problem was obtained \cite{DSW} by Damashke et al. in 2013. They provide an exact two stage group testing strategy and use $2.5\log_2t$ tests. For other constructions for the case of $2$ defects, we refer to \cite{mr98, DL}.  
	\section{Preliminaries}\label{Pre}
	
	Throughout the paper we use $t$, $s$, $p$ for the number of elements, defectives, and stages, respectively. 
	Let $\eq$ denote the equality by definition, $|A|$ -- the cardinality of the set $A$. The binary entropy function $h(x)$ is defined as usual $$h(x)=-x\log_2(x)-(1-x)\log_2(1-x).$$
	
	A binary $(N \times t)$-matrix with $N$ rows $\x_1, \dots, \x_N$ and $t$ columns $\x(1), \dots, \x(t)$ (codewords)
	$$
	X = \| x_i(j) \|, \quad x_i(j) = 0, 1, \quad i \in [N],\,j \in [t]
	$$
	is called a {\em binary code of length $N$  and size $t$}.
	The number of $1$'s in the codeword $x(j)$, i.e., $|\x(j)| \eq \sum\limits_{i = 1}^N \, x_i(j)= wN$,
	is called the {\em weight} of $\x(j)$, $j \in [t]$ and parameter $w$, $0<w<1$, is the \textit{relative weight}.
	
	One can see that the binary code $X$ can be associated with $N$ tests. A column $\x(j)$ corresponds to the $j$-th sample; a row $\x_i$ corresponds to the $i$-th test. 
	Let $\u \bigvee \v$ denote the disjunctive sum of binary columns $\u, \v \in \{0, 1\}^N$.
	For any subset $\S\subset[t]$ define the binary vector $$r(X,\S) = \bigvee\limits_{j\in\S}\x(j),$$
	which later will be called the \textit{outcome vector}.
	
	By $\S_{un}$, $|\S_{un}|\le s$, denote an unknown set of defects. Suppose there is a $p$-stage group testing strategy $\mathfrak{S}$ which finds up to $s$ defects. It means that for any $\S_{un}\subset[t]$, $|\S_{un}|\le s$, according to $\mathfrak{S}$:
	\begin{enumerate}
		\item we are given with a code $X_1$ assigned for the first stage of group testing;
		\item we can design a code $X_{i+1}$ for the $i$-th stage of group testing, based on  the outcome vectors of the previous stages $r(X_1,\S_{un})$, $r(X_2,\S_{un})$, \ldots, $r(X_i,\S_{un})$;
		\item we can identify all defects $\S_{un}$ using $r(X_1,\S_{un})$, $r(X_2,\S_{un})$, \ldots, $r(X_p,\S_{un})$.
		\end {enumerate}
		Let $N_i$ be the number of test used on the $i$-th stage and $$N_T(\mathfrak{S})=\sum_{i=1}^p N_i$$ be the maximal total number of tests used for the strategy $\mathfrak{S}$.
		We define $N_p(t, s)$ to be the minimal worst-case total number of tests needed for group
		testing for $t$ elements, up to $s$ defectives, and at most $p$ stages.
		
		\section{Hypergraph approach to searching defects}\label{Hyp}
		Let us introduce a hypergraph approach to searching defects. Suppose a set of vertices $V$ is associated with the set of samples $[t]$, i.e. $V = \{1,2\ldots, t\}$. 
		
		\textbf{First stage:}
		Let $X_1$ be the code corresponding to the first stage of group testing. For the outcome vector $r=r(X_1,\S_{un})$ let $E(r,s)$ be the set of subsets of $\S\subset V$ of size at most $s$ such that $r(X,\S)=r(X,\S_{un})$. So, the pair $(V,E(r,s))$ forms the hypergraph $H=H(X_1)$. We will call two vertices \textit{adjacent} if they are included in some hyperedge of $H$. Suppose there exist a \textit{good} vertex coloring of $H$ in $k$ colours, i.e., assignment of colours to vertices of $H$ such that no two adjacent vertices share the same colour. By $V_i\subset V$, $1\le i\le k$, denote vertices corresponding to the $i$-th colour. One can see that all these sets are pairwise disjoint. 
		
		\textbf{Second stage:}
		
		Now we can perform $k$ tests to check which of monochromatic sets $V_i$ contain a defect. Here we find the cardinality of set $\S_{un}$ and $|\S_{un}|$ sets $\{V_{i_1},\ldots,V_{i_{|\S_{un}|}} \}$, each of which contains exactly one defective element.
		
		\textbf{Third stage:}
		
		Carrying out $\lev\log_2|V_{i_1}|\riv$  tests we can find a vertex $v$, corresponding to the defect, in the suspicious set $V_{i_1}$. Observe that actually by performing $\sum\limits_{j=1}^{\S_{un}}\lev\log_2|V_{i_j}|\riv$ tests we could identify all defects $\S_{un}$ on this stage.
		
		\textbf{Fourth stage:}
		
		Consider all hyperedges $e\in E(r,s)$, such that $e$ contains the found vertex $v$ and consists of vertices of $v\cup V_{i_2}\cup \ldots\cup V_{i_{|\S_{un}|}}$. At this stage we know that the unknown set of defects coincides with one of this hyperedges. To check if the hyperedge $e$ is the set of defects we need to test the set $[t]\backslash e$. Hence, the number of test at fourth stage is equal to degree of the vertex $v$.
	
		\section{Optimal searching of 2 defects}\label{Search2}
		
		Now we consider a specific construction of $4$-stage group testing. Then we upper bound number of tests $N_i$ at each stage. 
		
		\textbf{First stage:}
		
		Let $C=\{0,1,\dots q-1\}^{\hat{N}}$  be the $q$-ary code, consisting of all $q$-ary words of length $\hat{N}$ and having size $t=q^{\hat{N}}$. 
		Let $D$ be the set of all  binary words with length $N'$ such that the weight of each codeword is fixed and equals $ wN'$, $0<w<1$, and the size of $D$  is at least $q$, i.e., $q \le {N' \choose {wN'}}$.
		On the first stage we use the concatenated binary code $X_1$ of length $N_1=\hat{N}\cdot N'$ and size $t=q^{\hat{N}}$, where the inner code is $D$, and the outer code is $C$. We will say $X_1$ consists of $\hat{N}$ layers.
		Observe that we can split up the outcome vector $r(X_1,\S_{un})$ into $\hat{N}$ subvectors of lengths $N'$. So let $r_j(X_1,\S_{un})$ correspond to $r(X_1,\S_{un})$ restricted to the $j$-th layer. Let $w_j$, $j\in[\hat{N}]$, be the relative weight of $r_j(X_1,\S_{un})$, i.e., $|r_j(X_1,\S_{un})| = w_jN'$ is the weight of the $j$-th subvector of $r(X_1,\S_{un})$.
		
		If $w_j=w$ for all $j\in[\hat{N}]$, then we can say that $\S_{un}$ consists of 1 element and easily find it.
		
		If there are at least two defects, then  suppose for simplicity that $\S_{un}=\{1,2\}$. The two corresponding codewords of $C$ are $c_1$ and $c_2$. There exists a coordinate $i, 1\le i\le \hat{N}$, in which they differs, i.e., $c_1(i)\neq c_2(i)$.  Notice that the relative weight $w_i$ is bigger than $w$. 
		
		For any $i\in[\hat{N}]$ such that $w_i>w$, we can colour all vertices $V$ in $q$ colours, where the colour of $j$-th vertex is determined by the 		
		corresponding $q$-nary symbol $c_i(j)$ of code $C$.
		 
		One can check that such a coloring is a good vertex coloring. 
		
		\textbf{Second stage:}
		
		We perform $q$ tests to find which coloured group contain $1$ defect.
		
		\textbf{Third stage:}
		
		Let us upper bound the size $\hat{t}$ of one of such suspicious group: 
		$$\hat{t}\le {w_1N' \choose wN'}\cdot\ldots \cdot {w_{\hat{N}}N' \choose wN'}.$$
		In order to find one defect in the group we may perform $\lev\log_2 \hat{t}\riv$ tests.
		
		\textbf{Fourth stage:}
		
		On the final step, we have to bound the degree of the found vertex $\it{v}\in V$ in the graph. The degree $\deg(\it{v})$ is bounded as
		$$
		\deg(\it{v})\le {wN' \choose (2w-w_1)N'}\cdot\ldots \cdot {wN' \choose (2w-w_{\hat{N}})N'}.
		$$
		We know that the second defect corresponds to one of the adjacent to $v$ vertices.
		Therefore, to identify it we have to make $\lev\log_2\deg(\it{v})\riv$ tests.
		
		The optimal choice of the parameter $w$ gives the procedure with total number of tests equals $2\log_2t(1+o(1))$.

	\end{document}